\documentclass[12pt, a4paper]{article}
\usepackage{amsfonts}
\usepackage{amssymb}
\usepackage{amsmath}
\usepackage{array}
\usepackage{bm}
\usepackage{cancel}
\usepackage{cite}
\usepackage{epsfig}
\usepackage{graphicx, rotating}
\usepackage[colorlinks=true,pdfstartview=FitV,linkcolor=blue,citecolor=blue,urlcolor=blue]{hyperref}
\usepackage{latexsym}
\usepackage{multirow}
\usepackage{slashed}
\usepackage{verbatim}
\usepackage{xcolor}
\usepackage[export]{adjustbox}
\usepackage[normalem]{ulem}
\usepackage[utf8]{inputenc}

\newcommand{\V}{{\mathcal V}}

\def\bma#1{\mbox{\boldmath{$#1$}}}

\numberwithin{equation}{section}

\begin{document}

\begin{titlepage}

\begin{flushright}
\small
DESY-26-057
\\
KOBE-COSMO-25-21
\end{flushright}
\vspace{.3in}

\begin{center}
{\Large\bf A positive definite formulation\\[0.2cm] of  vacuum decay with reduced symmetry}
\vskip 2mm
\bigskip\color{black}
\vspace{1cm}{
{\large
Jos\'e R.~Espinosa$^1$,
Ryusuke Jinno$^2$,
Thomas Konstandin$^3$,\\[0.2cm]
Shogo Matake$^2$, and Taiga Miyachi$^{4,5}$
}}

{\small
\vskip 7mm
$^1$ Instituto de F\'{\i}sica Te\'orica IFT-UAM/CSIC\\ 
C/ Nicol\'as Cabrera 13-15, Campus de Cantoblanco, 28049, Madrid, Spain\\
\vskip 1mm
$^2$ Department of Physics, Graduate School of Science, Kobe University, 1-1 Rokkodai, Kobe, Hyogo 657-8501, Japan\\
\vskip 1mm
$^3$ Deutsches Elektronen-Synchrotron DESY, Notkestr.~85, 22607 Hamburg, Germany\\
\vskip 1mm
$^4$ Department of Physics, Osaka Metropolitan University, Osaka 558-8585, Japan\\
$^5$ Osaka Central Advanced Mathematical Institute (OCAMI),
Osaka Metropolitan University, 3-3-138 Sugimoto, Sumiyoshi, Osaka 558-8585, Japan
}
%%%%%%%%%%%%%%%%%%%%%%%%%%%%%%%%%%%%%%%%%%%%%%%%%%

%%%%%%%%%%%%%%%%%%%%%%%%%%%%%%%%%%%%%%%%%%%%%%%%%%
%\date{\today}
\bigskip
\begin{abstract}
The Euclidean bounce for vacuum decay enjoys an $O(4)$ symmetry that is lost in the presence of impurities than can catalyze the decay.
We present a formulation for the calculation of the tunneling decay action, that is explicitly positive definite, for impurities whose effects are spherically symmetric so that the  bounce symmetry is reduced to $O(3)$. 
The action constructed can be regarded as a generalization of the  tunneling potential method, which implicitly assumed $O(4)$ symmetry.
We show that the action obtained reduces to the tunneling potential for $O(4)$-symmetric cases and provide analytic examples with $O(3)$ symmetry
and arbitrary wall thickness.
\end{abstract}
\end{center}
\end{titlepage}
%%%%%%%%%%%%%%%%%%%%%%%%%%%%%%%%%%%%%%%%%%%%%%%%%%

%%%%%%%%%%%%%%%%%%%%%%%%%%%%%%%%%%%%%%%%%%%%%%%%%%
\section{Introduction}
%%%%%%%%%%%%%%%%%%%%%%%%%%%%%%%%%%%%%%%%%%%%%%%%%%

Quantum tunneling is a truly significant phenomenon that appears in many physical systems. A particularly relevant instance is a first-order phase transition (FOPT), in which a scalar field initially trapped in a metastable (false) vacuum decays into a stable (true) vacuum. This phenomenon can occur in the early universe depending on the structure of the Higgs potential and may be probed through the observation of gravitational waves~\cite{LISACosmologyWorkingGroup:2022jok}.
The decay rate per unit volume controls the cosmological consequences after nucleation and is important in discussing the phenomenological consequences of vacuum decay.
The calculation of the decay rate is formulated in the seminal papers \cite{Coleman:1977py,Callan:1977pt,Coleman:1980aw,Coleman:1987rm} and the decay rate $\Gamma\sim\exp{(-S_E)}$ with the tunneling exponent given by the Euclidean action $S_E$ of the bounce solution, which is a solution to the classical equation of motion (EoM) with appropriate boundary conditions.

A novel formulation of the calculation of tunneling exponents that does not rely on Euclidean methods was proposed in Refs.~\cite{Espinosa:2018hue,Espinosa:2018voj}.
In this approach, the dynamical variable (called the tunneling potential) resides in field space rather than in position space. 
Among other attractive features,\footnote{The approach gives an intuitive picture of the tunneling; it is easy to approximate the action numerically \cite{Espinosa:2018hue}; it is ideally suited to discuss vacuum decay when there is no bounce solution~\cite{Espinosa:2019hbm} leading to the clearest understanding of pseudo-bounce decay \cite{Espinosa:2019hbm}; it provides a very clear picture of gravitational effects when generalized to nonzero gravity \cite{Espinosa:2018voj}, including bubble-of-nothing decays \cite{Blanco-Pillado:2023aom}.} the most notable property of this alternative formulation is the explicit positivity of the action.
Thanks to it we only need to search for the minimum of the action (rather than a saddle point as in the Euclidean formulation). This has immediate applications for the case of multi-field potentials which are numerically more tractable in this approach~\cite{Espinosa:2018szu}. 
As shown in~\cite{Espinosa:2022ofv} the tunneling potential action can be derived by a canonical transformation from the Euclidean action for
the $O(4)$-symmetric bounce and this powerful method can be generalized to cases with less symmetry, as we show in this paper.

Vacuum decay in the presence of impurities has attracted considerable attention as they can enhance the tunneling rate (for an incomplete list of references see ~\cite{Hiscook:1987,Gregory:2013hja,Mukaida:2017bgd,Kumar:2010mv,Agrawal:2022hnf,Oshita:2018ptr}) and they might exist in the universe.
One can have spherically symmetric objects such as monopoles~\cite{Dirac:1931kp,tHooft:1974kcl,Polyakov:1974ek,Kibble:1976sj,Zeldovich:1978wj,Preskill:1979zi}, I-balls~\cite{Bogolyubsky:1976yu,Copeland:1995fq,Gleiser:1993pt}, Q-balls~\cite{Coleman:1985ki}, and primordial black holes (PBHs)~\cite{Zeldovich:1967lct,Hawking:1971ei,Carr:1974nx,Chapline:1975ojl,Carr:1975qj}.
More generally, non-spherically symmetric impurities—such as spinning black holes \cite{Oshita:2019jan,Saito:2021vut}, cosmic strings with cylindrical symmetry and domain walls with planar symmetry~\cite{Kibble:1976sj}, can also catalyze vacuum decay. In general, such objects will 
break the $O(4)$ symmetry of the bounce solutions in vacuum (proven in \cite{Coleman:1977th}),
thereby necessitating a framework that does not rely on such symmetry. In this paper we will extend the tunneling potential formalism in this direction, focusing as a first step in the case with $O(3)$ symmetry: that is, the bounce is rotationally invariant in 3-space rather than in the full 4 dimensional
Euclidean space in which it lives. Even for $O(3)$-symmetric impurities, vacuum decay occurring off-center from a black hole can be significant \cite{Miyachi:2021bwd,Shkerin:2021zbf,Chang:2025rda}, and its configuration generally has less symmetry than $O(3)$. To keep things simple, in this paper we will consider the case in which the bounce has the defect at its core, so that the $O(3)$ symmetry is respected.

In less symmetric cases, it is very difficult to construct bounce solutions analogous to the $O(4)$ case, and consequently most works rely on the thin-wall approximation, discussed for the $O(4)$ case in \cite{Coleman:1977py}. However, the validity of the thin-wall approximation in less symmetric case is not guaranteed.
Therefore, for quantitative predictions of nucleation around such impurities, a symmetry-agnostic formulation is desirable for computing tunneling rates in non-symmetric configurations.

In this paper, we derive a generalized positive-definite tunneling action under the assumption of $O(3)$ symmetry only. The resulting tunneling action, given in Eq.~(\ref{eq:action_without_gravity}), is explicitly positive definite. While this property also holds in the $O(4)$-symmetric case, our formulation does not impose $O(4)$ symmetry a priori.  We demonstrate the applicability of the framework with some analytical examples, which are easy to construct.

The paper is organized as follows.
In Sec.~\ref{sec:review}, we first review the Euclidean formulation by Coleman and then explain how to derive the $O(4)$-symmetric positive-definite tunneling action via a canonical transformation.
In Sec.~\ref{sec:without_gravity}, we derive the generalized tunneling action for the $O(3)$ case. We also show how the generalized tunneling action has the appropriate $O(4)$ limit. In Sec.~\ref{sec:nonO(4)}, we discuss $O(3)$ examples that arise from deforming $O(4)$ solutions, presenting one analytic family of examples that includes both thick and thin-walled solutions. Sec.~\ref{sec:conclusion} is devoted to discussion and conclusions.

%%%%%%%%%%%%%%%%%%%%%%%%%%%%%%%%%%%%%%%%%%%%%%%%%%
\section{Review of the Euclidean method and the \texorpdfstring{$\bma{O(4)}$}{O(4)} tunneling action}
\label{sec:review}
%%%%%%%%%%%%%%%%%%%%%%%%%%%%%%%%%%%%%%%%%%%%%%%%%%

%%%%%%%%%%%%%%%%%%%%%%%%%%%%%%%%%%%%%%%%%%%%%%%%%%
\subsection{The Euclidean method}
\label{subsec:Euclidean}
%%%%%%%%%%%%%%%%%%%%%%%%%%%%%%%%%%%%%%%%%%%%%%%%%%

A key quantity related to tunneling in quantum field theory is the decay rate, which can be calculated with the Euclidean method developed in Refs.~\cite{Coleman:1987rm,Coleman:1977py,Callan:1977pt} for the case without gravity. 
In this subsection we briefly review how this method calculates the tunneling exponent.

Throughout this section, we consider the Euclidean action of a scalar field
\begin{align}
S_{E}
&=
\int d^4 x
\left[
\frac{1}{2} (\partial \phi)^2 + V (\phi)
\right]\,,
\label{eq:action_coleman}
\end{align}
where $V (\phi)$ is assumed to have a false vacuum at $\phi = \phi_f$ and a true vacuum at $\phi = \phi_t$.
The initial state of the scalar field in the false vacuum can tunnel through the potential barrier via quantum effects.
For sufficiently long-lived vacua the decay rate is exponentially suppressed as
\begin{align}
\Gamma\sim \exp{(-\bar{S}_E)}\,.
\label{eq:decayrate}
\end{align}
The tunneling exponent is the on-shell Euclidean action,
\begin{align}
\bar{S}_E = \int d^4\,x\Big[\frac{1}{2}(\partial \phi_B)^2+V(\phi_B)\Big]\,,
\end{align}
with $\phi_B$ being the bounce solution that obeys the classical EoM. 
In the absence of impurities, the bounce is $O(4)$-symmetric \cite{Coleman:1977th} and the EoM and the boundary conditions are 
\begin{align}
\ddot\phi_B+\frac{3}{r_E}\dot{\phi}_B&=V'(\phi_B)\,,
\\
\dot{\phi}_B(0)&=0\,,
\label{eq:shooting_initial}
\\
\phi_B(\infty)&=\phi_f\,,
\label{eq:shooting_BC}
\end{align}
where $r_E=\sqrt{x^2+y^2+z^2+\tau^2}$ denotes the radial coordinate in four-dimensional Euclidean space, $\phi_B(r_E)$ depends only on $r_E$ due to $O(4)$ symmetry, $\dot x\equiv dx/dr_E$ and $x'\equiv dx/d\phi$.
Condition (\ref{eq:shooting_initial}) comes from regularity at $r_E=0$ and condition (\ref{eq:shooting_BC}) comes from the physical assumption that $\phi$ is on the false vacuum side (initial state) at $\tau\to\pm\infty$.  

In the numerical calculations, the bounce solution is found by treating the free parameter $\phi(0) = \phi_0$ as a shooting parameter and adjusting it to satisfy the boundary condition at infinity. 
This commonly used procedure is referred to as the shooting method.

%%%%%%%%%%%%%%%%%%%%%%%%%%%%%%%%%%%%%%%%%%%%%%%%%%
\subsection{\texorpdfstring{$\bma{O(4)}$}{O(4)} tunneling potential action via a canonical transformation}
\label{subsec:O(4)}
%%%%%%%%%%%%%%%%%%%%%%%%%%%%%%%%%%%%%%%%%%%%%%%%%%

The tunneling potential method is a positive-definite formulation of the tunneling action which assumes $O(4)$ symmetry \cite{Espinosa:2018hue,Espinosa:2018voj,Espinosa:2022ofv}.
In this formulation, one only needs to find the configuration that minimizes the action because the solution is at a minimum point of positive-definite action rather than the saddle point of the Euclidean action.
Ref.~\cite{Espinosa:2022ofv} derived the positive-definite tunneling action via a canonical transformation. 
In this subsection, we briefly review how this was done. For a complete derivation following this method, we refer the reader to~\cite{Espinosa:2022ofv}.

We first consider the Euclidean action of a scalar field with $O(4)$ symmetry
\begin{align}
S_E=2\pi^2 \int_0^\infty  dr_E~r_E^3\Big[\frac{1}{2}\dot{\phi}^2+V(\phi)\Big]\,,
\end{align}
where the potential $V(\phi)$ is assumed to have a false vacuum at $\phi=\phi_f$ and a true vacuum at $\phi=\phi_t$.\\
We change the dynamical variable from $\phi(r_E)$ to its inverse $r_E(\phi)$, which is possible because the bounce is a monotonic function. Taking a $\phi$ derivative of the defining relation
$\phi(r_E(\phi))=\phi$, one gets 
\begin{align}
  \frac{\partial\phi}{\partial r_E}\frac{\partial r_E}{\partial\phi} = 1 
  \quad\implies\quad \dot{\phi} = \frac{1}{r_E'} \,,
\end{align}
(where we use a prime for derivatives with respect to $\phi$). 
We then rewrite the action with
$r_E(\phi)$ as the dynamical variable as
\begin{align}
S = 
\int_{\phi_f}^{\phi_0} d\phi \, L  = 
- 2\pi^2  \int_{\phi_f}^{\phi_0}  d\phi \; r_E^3 
   \left[ \frac{1}{2\,r_E'} + r_E'\,V \right]\,,
\label{eq:action_tp1}
\end{align}
with $\phi_0=\phi(r_E=0)$ and  $\phi_f=\phi(\infty)$. The overall minus sign arises from the reversal of the integration limits.

In this formulation, we interpret $\phi$ as the time and $r_E$ as a dynamical field. Starting from this Lagrangian, we perform a canonical transformation to obtain the desired action. The canonical momentum conjugate to $r_E$ is
\begin{align}
p &= \frac{\partial L}{\partial r_E'}
   = 2\pi^2 r_E^3 \left( \frac{1}{2r_E'{}^2} - V \right)
\quad\implies\quad
r_E' = - \left[2\!\left(V + \dfrac{p}{2\pi^2 r_E^3}\right)\right]^{-1/2} < 0 \,.
\end{align}
The derivative of $r_E$ with respect to $\phi$ is taken negative since $\phi(r_E)$ monotonically decreases in the Euclidean formulation.

To express the Hamiltonian in terms of the dynamical variables $(r_E,p)$, we perform a Legendre transformation and obtain
\begin{align}
H=p\, r_E'-L
=-2\pi^2r_E^3\sqrt{2\bigg(V+\frac{p}{2\pi^2r_E^3}\bigg)}\,.
\end{align}
In order to deform the action into the desired form, we perform a canonical transformation with generating function 
\begin{align}
G=-\frac{1}{2}\pi^2V_tr_E^4\,.
\end{align}
This canonical transformation changes the dynamical variables from $(r_E,p)$ to $(V_t,P)$. The new dynamical variable $V_t$ is called the tunneling potential.
Using the following relations,
\begin{align}
p&=\frac{\partial G}{\partial r_E}=-2\pi^2 r_E^3 V_t\,,
\qquad
P=-\frac{\partial G}{\partial V_t}=\frac{1}{2}\pi^2 r_E^4\,,
\label{eq:CT}
\end{align}
the new Hamiltonian is given by
\begin{align}
H'=H+\frac{\partial G}{\partial\phi}=-4(2\pi^2P^3)^{1/4}\sqrt{V-V_t}\,.
\label{eq:CT3}
\end{align}
Performing the Legendre transformation back to the Lagrangian form and then minimizing with respect to $P$, we obtain the desired action
\begin{align}
S[V_t]= 54\pi^2\int d\phi\frac{(V-V_t)^2}{(-V_t')^3}\,,
\label{eq:action_tunnelingpotential}
\end{align}
which is the one presented in \cite{Espinosa:2018hue}.

This action exhibits several features not present in the original formulation~(\ref{eq:action_coleman}). First, the action is positive definite, since $V_t'$ has negative value.
As noted at the beginning of this subsection, the positivity of the action significantly facilitates the search for the bounce solution, as the relevant solution can be obtained by minimizing a positive-definite functional rather than searching for a saddle point.
Second, the dynamical variable is given by  
\begin{align}
   V_t(\phi)=-\frac{p}{2\pi^2r_E^3}= V(\phi)-\frac{1}{2r_E'{}^2}=V(\phi)-\frac{1}{2}\dot{\phi}^2\,,
   \label{eq:Vt}
\end{align}
which is to be taken as a function of $\phi$, so that the entire  formulation is constructed in $\phi$-space, rather than in the $r_E$-space. That is, the formulation involves $V_t(\phi)$ rather than $\phi(r_E)$.

The EoM for $V_t$ is given by the Euler-Lagrange equation from (\ref{eq:action_tunnelingpotential}) as
\begin{align}
6 (V - V_t) V''_t + (4 V_t' - 3 V') V_t'
&=
0\,.
\label{eq:O(4)-tunneling_EoM}
\end{align}
The tunneling action $S[V_t]$, evaluated on the solution of this EoM, reproduces
the action obtained by solving for the bounce in the Euclidean approach. As an alternative, one may compute the on-shell action by minimizing the action with respect to $V_t$, e.g., using a gradient-flow method. 

In summary of this subsection, we started with the original action with $O(4)$ symmetry (\ref{eq:action_coleman}), changed the dynamical variable from $\phi(r_E)$ to $r_E(\phi)$ in (\ref{eq:action_tp1}), performed a canonical transformation to reformulate the Hamiltonian in (\ref{eq:CT}), and finally obtained the desired tunneling action (\ref{eq:action_tunnelingpotential}).

%%%%%%%%%%%%%%%%%%%%%%%%%%%%%%%%%%%%%%%%%%%%%%%%%%
\section{Generalized tunneling action}
\label{sec:without_gravity}
%%%%%%%%%%%%%%%%%%%%%%%%%%%%%%%%%%%%%%%%%%%%%%%%%%

The breaking of the $O(4)$ symmetry is due to some ambient impurity or defect to which the scalar field couples. We will model this via a position dependent scalar potential, $V(\phi,\vec{x})$, which is assumed to have terms that depend {\it both} on $\phi$ and $\vec{x}$.\footnote{Position dependent terms in $V$ that are independent of $\phi$ cannot affect the bounce as its EoM only depends on $\partial V/\partial\phi$.} For a spherically symmetric defect (like a monopole), we take $V=V(\phi,r)$, where $r$ is the radial distance to the center of the bounce, which we expect to have $O(3)$ symmetry, $\phi_B(r,\tau)$. As already explained in the introduction, this assumes that the defect sits at the center of the bounce.
When the impurity favors a tunneling event in its neighborhood this assumptions 
is very plausible. Ultimately, it will depend on the potential $V(\phi,r)$
if the most likely tunneling configuration is away from the impurity,
close to it or even concentric with the impurity, as we assume here.

%%%%%%%%%%%%%%%%%%%%%%%%%%%%%%%%%%%%%%%%%%%%%%%%%%
\subsection{Derivation}
\label{subsec:without_gravity_derivation}
%%%%%%%%%%%%%%%%%%%%%%%%%%%%%%%%%%%%%%%%%%%%%%%%%%

In this section we present the derivation of a positive-definite action of tunneling for an $O(3)$-symmetric bounce. 
The positive-definite tunneling action is derived via a canonical transformation, analogous to the approach used in the previous section for the 
$O(4)$ case, but only assuming $O(3)$ symmetry.

We start with the Euclidean action of a scalar field  with $O(3)$ symmetry,
\begin{align}
S_{E}=\int d\tau\int 4\pi r^2 dr\Bigg[\frac{1}{2}\left(\frac{\partial\phi}{\partial \tau}\right)^2+\frac{1}{2}\left(\frac{\partial\phi}{\partial r}\right)^2+V(\phi,r)\Bigg].
\end{align}
Here $\phi(\tau,r)$ is the dynamical variable and $(\tau,r)$ are the independent variables. The potential is assumed to have a false vacuum at $\phi_f(r)$ and a deeper (true) vacuum at $\phi_t(r)$. Notice that the $r$-dependence of $V$ implies that these vacua can be non-homogeneous stationary solutions of
\begin{align}
    \frac{\partial^2\phi}{\partial r^2}+\frac{2}{r}\frac{\partial\phi}{\partial r}=\frac{\partial V}{\partial \phi}\ .
\end{align}
However this is not necessarily the case: in the concrete example we show in section~\ref{subsec:thinwallO(3)}, false and true vacua are both homogeneous.

The bounce solution satisfies the following boundary conditions
\begin{align}
\phi_B(\tau,\infty)=\phi_f(\infty),\quad\quad
\phi_B(\pm\infty,r)=\phi_f(r),\quad\quad
\left.\frac{\partial\phi_B}{\partial r}\right|_{r=0}=0,
\end{align}
which follow from the finiteness of the action and regularity at $r=0$.

To obtain a generalized positive-definite action density, we change the dynamical variables from $\phi(\tau,r)$ to $\tau(\phi,r)$.
Taking $\phi$ and $r$ derivatives of the defining relation $\phi(\tau(\phi,r),r)=\phi$, one gets
%%%
\begin{align}
 \quad \frac{\partial \phi}{\partial \tau}\frac{\partial \tau}{\partial \phi}=1 \quad &\implies\quad \frac{\partial\phi}{\partial\tau}=\frac{1}{\tau'}\,,\\
\frac{\partial\phi}{\partial \tau}\frac{\partial \tau}{\partial r}+\frac{\partial\phi}{\partial r}=0 \quad &\implies\quad
 \tau'\,\dot{\phi}=- \dot{\tau}\,,
\end{align}
as well as
\begin{align}
    \frac{\partial^2\phi}{\partial\tau^2}=-\frac{\tau''}{\tau'{}^3}\ ,\quad
    \frac{\partial^2 \phi}{\partial r\partial\tau}=-\frac{\dot{\tau}'}{\tau'{}^2}+\frac{\tau''\dot{\tau}}{\tau'{}^3} \ ,\quad
    \frac{\partial^2\phi}{\partial r^2}=-\frac{\ddot{\tau}}{\tau'}-\frac{\tau''\dot{\tau}^2}{\tau'{}^3}+\frac{2\dot{\tau}\dot{\tau}'}{\tau'{}^2}\ .
\end{align}
where prime and dot denote $\phi$ and $r$ derivatives, respectively.
In these variables the EoM for the bounce reads
\begin{align}
    -\frac{\tau''}{\tau'{}^3}\left(1+\dot{\tau}^2\right) -\frac{1}{\tau'}\left(\ddot{\tau}+\frac{2}{r}\dot{\tau}\right)+\frac{2\dot{\tau}\dot{\tau}'}{\tau'^2}=V'\ .
\label{EoMtau}
\end{align}

The transformed action is
\begin{align}
S
=\int d\phi\int dr\, {\cal L} 
=
\int d\phi\int4\pi r^2 dr\Bigl\lbrack\frac{1+\dot{\tau}^2}{2\tau'}+\tau'V(\phi,r)\Bigr\rbrack,
\label{eq:action_tau}
\end{align}
where the dynamical variable is $\tau(\phi,r)$.

The canonical momentum density conjugate to $\tau'$, called $p$, is 
\begin{align}
p=\frac{\partial \mathcal{L}}{\partial \tau'}=4\pi r^2\Bigl\lbrack-\frac{1+\dot{\tau}^2}{2\tau'^2}+V(\phi,r)\Bigr\rbrack 
\;\Longrightarrow\; \tau'=\sqrt{\frac{1+\dot{\tau}^2}{2\left[V-p/(4\pi r^2)\right]}}>0.\label{eq:conjugate_momentum}
\end{align}
The sign of $\tau'$ corresponds to the region $\tau<0$, assuming $\tau'$ to be positive since $\partial\phi/\partial\tau$ is assumed to be positive. 
This region corresponds to half of the entire Euclidean spacetime, and we later multiply the resulting positive-definite action by 2 to account for the full spacetime.   Notice that, at fixed $r$, $\phi$ covers then only one half of the bounce trajectory in Euclidean time. It starts from the false-vacuum side and ends at the turning point in the $\tau$ direction, where $\partial_\tau\phi=0$. By the $\tau\to-\tau$ symmetry of the bounce, this point is located at $\tau=0$. Thus the endpoint of the $\phi$-integration is identified with the $\tau=0$ slice of the bounce, $\phi_B(0,r)$.

The Hamiltonian density is obtained via a Legendre transformation as
\begin{align}
\mathcal{H}=p\tau'-\mathcal{L}=-4\pi r^2\sqrt{2(1+\dot{\tau}^2)\left[V-p/(4\pi r^2)\right]}
\label{eq:Hamiltonian_WOO(4)}
\end{align}
Now we perform a canonical transformation to change the variables from $(\tau,p)$ to $(Q,P)$. We take the generating function to be 
\begin{align}
W[\tau,P,\phi] &= \int dr~\tau\dot{P},\label{eq:generating_function}
\end{align}
which yields the following relations
\begin{align}
p &= \frac{\delta W}{\delta \tau}=\dot{P},
\label{pPdot}\\
Q &= \frac{\delta W}{\delta P}=-\dot{\tau}.
\label{Qtaudot}
\end{align}
Here the functional derivative with respect to $P$ should be understood as taken 
after integrating by parts with respect to $r$. More explicitly, the variation
of $W$ is
\begin{align}
\delta W
=
\int dr \left(\dot P\,\delta\tau-\dot\tau\,\delta P\right)
+\left.\left(\tau\,\delta P\right)\right|_{0}^{r_M}\,, 
\end{align}
where $r_M$ is the upper boundary of integration for $r$ at fixed $\phi$.
The functional derivative is therefore taken for variations within a fixed
 boundary sector, so that $\delta P$ vanishes at the spatial boundaries.
With this boundary condition, the last term does not contribute, and we obtain  Eqs.~\eqref{pPdot} and \eqref{Qtaudot}.
The new Hamiltonian is then
\begin{align}
\widetilde{H}
&=
H+\frac{\partial W}{\partial\phi}=\int \widetilde{\mathcal H}\, dr=-\int dr~4\pi r^2 \sqrt{2(1+Q^2)\left[V-\dot{P}/(4\pi r^2)\right]}\,,
\end{align}
where $\partial W/\partial\phi=0$ since the generating functional has no explicit dependence on $\phi$. 
The corresponding  action is obtained by a Legendre transformation as
\begin{align}
S
=&
\int d\phi\int dr \left.\left( p\tau'-\widetilde{\mathcal{H}}\right)\right|_{Q}
\notag\\
=&
\int d\phi\int dr \left.\left\{
\dot{P}\tau'
+4\pi r^2 \sqrt{2(1+\dot{\tau}^2)[V-\dot{P}/(4\pi r^2)]}
\right\}\right|_{Q}.
\label{eq:SE_Legendre}
\end{align}
We now integrate the first term by parts with respect to $r$. Since the
integration region in the $(\phi,r)$ plane has $r$-dependent boundaries\footnote{We show an explicit example of this in Subsec.~\ref{subsec:thinwallO(3)}.}
$\phi=\phi_m(r)$ and $\phi=\phi_M(r)$ (we expect $\phi_M(r)=\phi_B(r,0)$ as explained above), this gives
\begin{align}
S
=&
\int d\phi\int dr \left.\left\{
-P \dot{\tau'}
+4\pi r^2 \sqrt{2(1+\dot{\tau}^2)[V-\dot{P}/(4\pi r^2)]}
\right\}\right|_{Q}
\notag\\
&+\left.\int_{\phi_m(r)}^{\phi_M(r)} d\phi\left(P\tau'\right)\right|^{r=\infty}_{r=0}
-\int dr \left.\left(P\tau'\frac{d\phi_B}{dr}\right)\right|^{\phi_{M}(r)}_{\phi_{m}(r)}.
\label{eq:SE_after_r_ibp}
\end{align}
Using $Q=-\dot{\tau}$, the first term in the bulk part becomes $P Q'$.
A further integration by parts with respect to $\phi$ then gives
\begin{align}
S
=&
\int d\phi\int dr \left.\left\{
-P' Q
+4\pi r^2 \sqrt{2(1+Q^2)[V-\dot{P}/(4\pi r^2)]}
\right\}\right|_{Q}
\notag\\
&- \int dr\left.\left(P\dot{\tau}\right)\right|^{\phi_{M}(r)}_{\phi_{m}(r)}
+\left.\int_{\phi_m(r)}^{\phi_M(r)} d\phi\left(P\tau'\right)\right|^{r=\infty}_{r=0}
-\int dr \left.\left(P\tau'\frac{d\phi_B}{dr}\right)\right|^{\phi_{M}(r)}_{\phi_{m}(r)}.
\label{eq:SE_after_phi_ibp}
\end{align}
The boundary terms come from the integration by parts with respect to $r$ and $\phi$.
Note that integration by parts with respect to the spatial coordinates hits
$\phi=\phi_m(r)$ and $\phi=\phi_M(r)$ as well, which yields the last term.
In the following calculation, the first and last boundary terms cancel each other because
\begin{align}
&\int dr\left.\left(P\dot{\tau}\right)\right|^{\phi_{M}(r)}_{\phi_{m}(r)}+\int dr \left.\left(P\tau'\frac{d\phi_B}{dr}\right)\right|^{\phi_{M}(r)}_{\phi_{m}(r)}\notag
\\=&\int dr\left.\left(P\dot{\tau}\right)\right|^{\phi_{M}(r)}_{\phi_{m}(r)}+\int dr \left.\left[P\tau'\Big(-\frac{\dot{\tau}}{\tau'}\Big)\right]\right|^{\phi_{M}(r)}_{\phi_{m}(r)}=0,
\end{align}
where we have used $d\phi_B/dr=-\dot{\tau}/\tau'$.
By minimizing the action with respect to $Q$, we find the following relations from the bulk action, 
\begin{align}
P'-4\pi r^2\sqrt{2\left[V-\dot{P}/(4\pi r^2)\right]} \frac{Q}{\sqrt{1+Q^2}}=0
\notag\\\Longrightarrow 
Q=\frac{P'}{\sqrt{32\pi^2 r^4\Big(V-\frac{\dot{P}}{4\pi r^2}-\frac{P'^2}{32\pi^2r^4}\Big)}}.\label{eq:def_Q}
\end{align}
The remaining surface term can be expressed in terms of $P$. From (\ref{eq:conjugate_momentum}) and (\ref{eq:def_Q}), $\tau'$ is given by
\begin{align}
\tau'=\frac{1}{\sqrt{2\Big(V-\frac{\dot{P}}{4\pi r^2}-\frac{P'^2}{32\pi^2r^4}\Big)}}
\end{align}
Substituting this and multiplying by a factor of two to account for the other half of the bounce, we obtain the positive-definite tunneling action,
\begin{align}
S=\int d\phi\int dr\sqrt{128\pi^2 r^4 \left(V-\frac{\dot{P}}{4\pi r^2}-\frac{P'^2}{32\pi^2 r^4}\right)}+ \delta_B S\ .
\label{SP}
\end{align}
For ordinary finite-action bounces, $\delta_BS$ is zero, as is discussed in the next subsection. 

We now define the tunneling potential through
\begin{align}
V_t
&=
\frac{3}{4 \pi r^3} P.
\label{VtP}
\end{align}
To make contact with the Euclidean quantities and the $O(4)$ definition of $V_t$, notice that (\ref{eq:conjugate_momentum})
can be written as 
\begin{align}
    \frac{p}{4\pi r^2} = V(\phi,r) - \frac12 \dot\phi_B^2 - \frac12 (\partial_\tau\phi_B)^2 \equiv {\cal V}_t(\phi,r)\ .
\end{align}
Using also (\ref{pPdot}), we get
\begin{align}
    V_t(\phi,r)=\frac{3}{r^3}\int^r  \bar r{}^2 {\cal V}_t(\phi,\bar r)d\bar r\ .
    \label{link}
\end{align}
We recover the $O(4)$ definition of $V_t=V-(d\phi/dr_E)^2/2$ when ${\cal V}_t$ is $r$-independent.

Using (\ref{VtP}) we finally arrive at the positive-definite action with O(3)-symmetry expressed in terms of the tunneling potential $V_t$,
\begin{align}
S[V_t]=
\int d\phi \int dr~
\sqrt{128 \pi^2 r^4 \left[ V - \left( V_t + \frac{r}{3} \dot{V}_t + \frac{r^2}{18} {V'_t}^2 \right) \right]}\ ,
\label{eq:action_without_gravity}
\end{align}
where we have already dropped the zero boundary term.
This is a generalized tunneling action of $O(4)$ tunneling potential (\ref{eq:action_tunnelingpotential}) with only $O(3)$ symmetry.
Because the derivation is based solely on a canonical transformation, the resulting action yields the same on-shell value as the original bounce formulation. Indeed, a canonical transformation preserves the on-shell action by definition.

It is clear that the action density is positive definite since it is written as a square root of a positive quantity as in the $O(4)$ tunneling potential method. In contrast to the $O(4)$-symmetric tunneling potential, which depends only on $\phi$, the generalized tunneling potential depends on both $\phi$ and $r$. This reflects the fact that the original $O(3)$-symmetric scalar field is $\phi(\tau,r)$, and that we have interchanged the roles of $\phi$ and $\tau$. After this change of variables, the radial coordinate $r$ remains as an independent coordinate, so the tunneling potential naturally acquires an explicit $r$-dependence. In Sec.~\ref{sec:nonO(4)}, we will present an $r$-dependent solution based on the analytic deformed Fubini solution. Furthermore, $V_t(\phi,r)$ is related to the $O(4)$-symmetric tunneling potential $V_t(\phi)$. As we show in Subsec.~\ref{subsec:without_gravity_O(4)limit}, by taking an appropriate limit, the action reduces to the $O(4)$-symmetric tunneling potential action (\ref{eq:action_tunnelingpotential}). 

We integrate over the two-dimensional region in $(\phi,r)$-space where the quantity inside the square root is not negative as the action should be a real number. With this choice of integration range, the locus where the quantity under the square root vanishes gives the $\tau=0$ slice of the bounce in the original formulation, $\phi_B(0,r)$. This follows from the assumption that $\phi_B(\tau,r)$ decreases monotonically in the \(\tau\) direction, for $\tau>0$, so that for fixed $r$, the upper endpoint $\phi_M(r)$ of the $\phi$-integration is identified with $\phi_B(0,r)$. So, depending on the order of integration we can write such integrals in the two alternative forms
\begin{align}
    \int_0^\infty dr \int_{\phi_f(r)}^{\phi_B(0,r)} d\phi\, f(r,\phi) = \int_{\phi_m}^{\phi_0}d\phi \int_{r_m(\phi)}^{r_M(\phi)} dr\, f(r,\phi) \ . 
\end{align}
In the LHS integral, the full range of $r$ is of course $(0,\infty)$ while for fixed $r$ the field ranges from $\phi_f(r)$ to $\phi_B(0,r)$.
In the RHS integral instead, $\phi_m$ is the smallest value taken by the field, while the maximum value is assumed to be $\phi_0=\phi(0,0)$. For the limits of the $r$ integral we have $r_{m,M}(\phi)$ defined by  $\phi_f(r_m(\phi))=\phi$ and $\phi_B(0,r_M(\phi))=\phi$, where $\phi_f(r)$ is the false vacuum.
For a concrete example of such integration region see Fig.~\ref{fig:Stw} in Sec.~\ref{subsec:thinwallO(3)}. 

The EoM is easily derived as\footnote{Using (\ref{link}), this EoM for $V_t$ reduces to the Euclidean EoM.}
\begin{align}
6 (V - V_t) V_t'' + (4V_t' - 3 V') V_t' + 3 \ddot{V}_t + 2 r (V_t' \dot{V}_t' - V_t'' \dot{V}_t) + \frac{3}{r}(4\dot{V}_t - 3 \dot{V})
&=
0\,.
\label{eq:EOMVt}
\end{align}
Since the dynamical variable $V_t(\phi,r)$ depends on both $\phi$ and $r$, the resulting EoM is a partial differential equation rather than an ordinary differential equation. In the appropriate limit, this equation also reduces to the EoM for the $O(4)$-symmetric tunneling potential, see Subsec.~\ref{subsec:without_gravity_O(4)limit}.

%%%%%%%%%%%%%%%%%%%%%%%%%%%%%%%%%%%%%%%%%%%%%%%%%%
\subsection{Boundary term of the action
\label{subsec:deltaBS}}
%%%%%%%%%%%%%%%%%%%%%%%%%%%%%%%%%%%%%%%%%%%%%%%%%%

The boundary term of the action in (\ref{SP}) is given by
\begin{align}
\delta_B S=\left.\int_{\phi_{\rm m}(r)}^{\phi_{\rm M}(r)} d\phi~
2P\left[2 \left (V - \frac{\dot{P}}{4\pi r^2} + \frac{P'^2}{32\pi r^4}\right) \right]^{-1/2}
\right|^{r=\infty}_{r=0},
\end{align}
and in terms of $V_t$ this reads
\begin{align}
\delta_B S=
\left.\int_{\phi_{\rm m}(r)}^{\phi_{\rm M}(r)} d\phi~
8\pi r^3\,V_t\left\{18 \left[ V - \left( V_t + \frac{r}{3} \dot{V}_t + \frac{r^2}{18} V_t'{}^2 \right) \right]\right\}^{-1/2}
\right|^{r=\infty}_{r=0}.\label{eq:action_without_gravity_bound}
\end{align}
For finite-action bounces, the boundary term from spatial infinity is zero.
Indeed, a finite-action bounce approaches the false vacuum at spatial infinity. 
With the normalization $V(\phi_f,r)=V_t(\phi_f,r)=0$, the integration field interval in this surface term shrinks to zero as
$\phi_{\rm M}(r)\to \phi_f(r)$ for $r\to\infty$.
Moreover, for an ordinary potential with a false vacuum with nonzero positive mass for $\phi$, this approach is exponential, up to powers of $r$. 
The surface term contains only polynomial powers of $r$, while the endpoint singularity from the square root is integrable. 
Therefore the shrinking of the integration range in $\phi$ suppresses the surface contribution, and the boundary term vanishes.

In other cases of interest, like $V(\phi)=-\lambda\phi^4/4$ which has a vacuum at $\phi=0$ with zero mass, the bounce has a power-law tail rather than an exponential tail. In this case, with a Fubini-type solution deformed in the $r$ direction, one can show from the asymptotic behavior at infinity that the boundary term also vanishes, provided that the bulk action is finite. 

On the other hand, the contribution from $r=0$ vanishes for regular configurations. Since $P=4\pi r^3 V_t/3$, the $r=0$ boundary integrand is proportional to $r^3 V_t \tau'$. For a regular bounce and a regular potential, $V_t$ is finite at the origin. Moreover, $\tau'=1/\partial_\tau\phi_B$ at fixed $r$, so a possible divergence of $\tau'$ can occur only where $\partial_\tau\phi_B=0$, e.g. at the turning point $\phi=\phi_B(0,r)$, where $\tau'$ has the usual integrable square-root singularity in the $\phi$-integration. Therefore the $\phi$-integral is finite, and the explicit factor of $r^3$ makes the $r=0$ boundary contribution vanish.

%%%%%%%%%%%%%%%%%%%%%%%%%%%%%%%%%%%%%%%%%%%%%%%%%%
\subsection{\texorpdfstring{$\bma{O(4)}$}{O(4)} limit}
\label{subsec:without_gravity_O(4)limit}
%%%%%%%%%%%%%%%%%%%%%%%%%%%%%%%%%%%%%%%%%%%%%%%%%%

Now we show that the tunneling action (\ref{eq:action_without_gravity}) reduces to the correct $O(4)$-symmetric tunneling potential action 
(\ref{eq:action_tunnelingpotential}) in the $O(4)$ limit. In that limit\footnote{
An $O(4)$-symmetric bounce depends only on $r_E=\sqrt{r^2+\tau^2}$.
Then fixing $\phi$ is equivalent to fixing $r_E$, and thus we have
$\partial V_t/\partial r\big|_\phi = \partial V_t/\partial r\big|_{r_E}
$.
The RHS vanishes because an $O(4)$-symmetric quantity is constant along surfaces of fixed $r_E$.
Therefore, $V_t$ with $O(4)$ symmetry has no $r$-dependence.} $V_t$ is independent of $r$ and 
 we can carry out the integration over $r$ in the tunneling action (\ref{eq:action_without_gravity}) that takes the form
\begin{align}
S [V_t]
&=
\int^{\phi_B(0,r)}_0 d\phi \int_0^{r_{\rm M}} dr~
\sqrt{128 \pi^2 r^4 \left[ V - V_t - \frac{r^2}{18} V_t'{}^2 \right]}\,.
\end{align}
The upper limit of the $\phi$ integration at fixed $r$ is the nucleated bubble configuration $\phi_M(r)=\phi_B(0,r)$ and the lower 
one is the false vacuum which we take to be $\phi=0$.
The upper limit of the $r$ integration $r_{\rm M}$ can be determined from the condition that the quantity inside the square root be non-negative. Thus
\begin{align}
r_{\rm M}^2
&=
\frac{18 (V - V_t)}{V_t'{}^2}\,.
\end{align}
Notice that this is precisely the standard $O(4)$ relation between the Euclidean radial coordinate and $V_t$ \cite{Espinosa:2018hue}.
Changing the integration variable from $r$ to $r^2$ we get
\begin{align}
S [V_t]=
\int^{\phi_B(0,r)}_0 d\phi \int_0^{r_{\rm M}^2} d(r^2)~
\frac{4 \pi}{3} (- V_t') \sqrt{ \frac{18 (V - V_t)}{V_t'{}^2} r^2- r^4 }\, .
\end{align}
Using $\phi_B(0,\infty)=0$ and $\phi_B(0,0)=\phi_0$, we finally get
\begin{align}
S [V_t]=54 \pi^2 
\int_0^{\phi_0} d\phi~
\frac{(V - V_t)^2}{(- V_t')^3}\,,
\label{eq:O(4)limit}
\end{align}
which is precisely the $O(4)$-symmetric tunneling potential action \eqref{eq:action_tunnelingpotential}.

Obviously, the EoM (\ref{eq:EOMVt}) also has the correct $O(4)$ limit and for $V_t$ independent of $r$ it reduces to
\begin{align}
6 (V - V_t) V''_t + (4 V_t' - 3 V') V_t'
&=
0\,,
\end{align}
which is precisely the known EoM for the $O(4)$-symmetric tunneling potential. 

%%%%%%%%%%%%%%%%%%%%%%%%%%%%%%%%%%%%%%%%%%%%%%%%%%
\section{Some analytic solutions with \texorpdfstring{$\bma{O(3)}$}{O(3)} symmetry}
\label{sec:nonO(4)}
%%%%%%%%%%%%%%%%%%%%%%%%%%%%%%%%%%%%%%%%%%%%%%%%%%

In order to get analytic bounce solutions with only $O(3)$ symmetry (say due to some point-like defect), 
we will deform a given $O(4)$ analytic bounce using some simple deformation function. For instance, we can consider the deformation
\begin{align}
\phi_B(\sqrt{r^2+\tau^2}) \to \phi_B(\sqrt{a^2(r)r^2+\tau^2})\ ,
\end{align}
where $r$ is the radial distance to the center of the bounce, which is $O(3)$ symmetric, and $a(r)$ is the deformation 
function.\footnote{We leave for future work constructing in this way bounce solutions with even less symmetry.}
Obviously, this is just a particularly simple class of possible $O(3)$-symmetric solutions. 

Any acceptable deformation with $a\neq 1$ should have some properties. In particular, 
we need that the 3-dimensional nucleated bubble, $\phi_B$ at $\tau=0$, has zero energy (taken to be the energy of the false vacuum). 
We will also require that at spacial infinity far from the defect ($r\to\infty$) we fall back to the false vacuum 
and with $a\to 1$, so that the deforming influence of the defect dies off sufficiently fast.

Thus, our starting point is an $O(4)$ vacuum decay solution for some potential $V_0(\phi)$. In the Euclidean
formulation the solution takes the form of a bounce $\phi_{B0}(r_E)$, where $r_E$ stands for the Euclidean radial 
distance, while in the tunneling potential formulation the solution is a function $V_{t0}(\phi)$. 
The usual relations linking both formalisms hold. For later use we collect them here. The key relation is
\begin{align}
    V_{t0}(\phi)=V_0(\phi)-\frac12 \dot{\phi}_{B0}^2\ .
\end{align}
In this equation, the bounce slope is expressed in terms of the field $\phi$ using the inverted bounce profile
function $r_{E\phi}(\phi)$, defined by
\begin{align}
    \phi_{B0}(r_{E\phi}(\phi))=\phi\ .
    \label{rE}
\end{align}
We also define 
\begin{align}
    r_{Ea}^2\equiv a^2r^2+\tau^2\ ,
\end{align}
so that
\begin{align}
    r_{E\phi}^2(\phi_B)=r_{Ea}^2\ .
\end{align}

The EoM for $V_{t0}$ is
\begin{align}
    (4V_{t0}'-3V_0')V_{t0}'+6(V_0-V_{t0})V_{t0}''=0\ ,
    \label{EoMVt0}
\end{align}
and, from the EoM for $\phi_{B0}$, we get
\begin{align}
    r_{E\phi}^2(\phi)=\frac{18(V_0-V_{t0})}{V_{t0}'{}^2}\ .
    \label{rE2}
\end{align}
The following relation, which follows from (\ref{EoMVt0}) and (\ref{rE2}), will be of use later on
\begin{align}
    r_{E\phi}^2{}'=\frac{6}{V_{t0}'}\ .
    \label{rE2p}
\end{align}

In the next subsection we discuss general features of this class of deformed solutions, obtaining the 
position-dependent potentials that would give rise to such bounces and simple formulas for their tunneling
action  as well as for the energy of the nucleated bubbles for vacuum decay. After that subsection we present 
one concrete example of tunneling solution with $O(3)$ symmetry just using an existing analytical $O(4)$ 
bounce.\footnote{As it usually happens, the simple relations we obtain in Section~\ref{subsec:general} for the general case
were derived  only after they were found to hold in several analytical cases. This is just another instance of the usefulness of
analytically controlled examples.}

%%%%%%%%%%%%%%%%%%%%%%%%%%%%%%%%%%%%%%%%%%%%%%%%%
\subsection{Some general results for bounce deformations}
\label{subsec:general}
%%%%%%%%%%%%%%%%%%%%%%%%%%%%%%%%%%%%%%%%%%%%%%%%%

Our deformed bounce solution is \begin{align}
    \phi_B(\vec{x},\tau)=\phi_{B0}(\sqrt{a^2r^2+\tau^2})\ .
\end{align}
We can exchange $\tau$ by $\phi$ using the relation 
\begin{align}
    \tau^2=r_{E\phi}^2(\phi_B)-a^2r^2\ .
    \label{tauphi}
\end{align}
We introduce a vector $\vec{w}_t$ related to the gradient of the deformed bounce by
\begin{align}
    \nabla \phi_B = \frac{\nabla r_{Ea}}{r_{E\phi}'(\phi_B)}\equiv \frac{2 \vec{w}_t}{r_{E\phi}^2{}'(\phi_B)} = \frac13 V_{t0}'(\phi_B) \vec{w}_t \ .
    \label{wt}
\end{align}
We have
\begin{align}
\vec{w}_t=\frac12 \nabla (a^2r^2)=
 r a(r)\left[a(r)+r\dot a(r)\right] \vec{e}_r\ ,
\end{align}
where $\vec{e}_r$ is the unit vector along the radial direction in spherical coordinates. In the $O(4)$ symmetric limit $a=1$, we obtain 
$\vec{w_t}=r\, \vec{e}_r$. 

Then, we also have
\begin{align}
    \nabla^2\phi_B = \frac13 V_{t0}'(\phi_B) \nabla\cdot\vec{w}_t + \frac19 V_{t0}'(\phi_B) V_{t0}''(\phi_B)  \vec{w}_t^2\, ,
\end{align}
with
\begin{align}
\nabla\cdot\vec{w}_t=
 \frac{1}{r^2}\frac{d}{dr}\left\{r^3 a(r)\left[a(r)+r\dot a(r)\right] \right\}\, .
\end{align}
In the $O(4)$ symmetric limit $a=1$, $\nabla\cdot\vec{w}_t=3$. 

For the $\tau$ derivatives we have
\begin{align}
    \frac{\partial\phi_B}{\partial \tau} = \frac{2\sqrt{r_{Ea}^2-a^2r^2}}{r_{E\phi}^2{}'(\phi_B)} = \frac13 V_{t0}'(\phi_B)\sqrt{r_{Ea}^2-a^2r^2}\ .
    \label{dtaudphi}
\end{align}
and
\begin{align}
    \frac{\partial^2\phi_B}{\partial \tau^2} = V_0'(\phi_B)-V_{t0}'(\phi_B)-\frac19 a^2r^2 V_{t0}'(\phi_B)V_{t0}''(\phi_B)\ .
    \label{dtau2dphi}
\end{align}

Using the previous results, the bounce EoM
\begin{align}
    \frac{\partial^2\phi_B}{\partial \tau^2}+\nabla^2\phi_B
    =\frac{\partial V}{\partial \phi}(\phi_B,\vec{x})\ ,
\end{align}
can be integrated in $\phi_B$ to get the potential, which we then write in terms of a generic field $\phi$ as
\begin{align}
    V(\phi,\vec{x})=V_0(\phi)-V_{t0}(\phi)
    +\frac13 V_{t0}(\phi)\, \nabla\cdot\vec{w}_t
    +\frac{V_{t0}'(\phi)^2}{18}(\vec{w}_t^2-a^2r^2)\ .
\label{Vphix}
\end{align}
In the $O(4)$ symmetric limit we simply obtain $V=V_0(\phi)$ as we should.

Using the previous results, the "tunneling potential density" ${\cal V}_t$ is given by the simple relation
\begin{align}
    \V_t(\phi,\vec{x}) \equiv V(\phi,\vec{x})-\frac12 \left(\frac{\partial \phi_B}{\partial \tau}\right)^2 -\frac12 \left(\nabla \phi_B\right)^2
    =\frac{V_{t0}(\phi)}{3}\nabla\cdot \vec{w}_t\ .
    \label{Vtwt}
\end{align}
The tunneling potential is then 
\begin{align}
    V_t(\phi,\vec{x})=V_{t0}(\phi)a(r)\left[a(r)+r \dot a(r)\right]\, .
\end{align}
It can be checked that the EoM  (\ref{eq:EOMVt}) is satisfied.

For the action we get
\begin{align}
    S&=\frac{2}{3}\int_0^{\phi_B(\vec{x},0)}d\phi\,
    \int d^3x\,  
    \sqrt{\left[r_{E\phi}^2(\phi)-a^2r^2\right]V'_{t0}{}^2(\phi)}\nonumber\\
    &+ 2 \int_0^{\phi_B(\vec{x},0)}d\phi\,
     \int d\vec{S} \cdot  
    \frac{V_{t0}(\phi)\vec{w_t} }{\sqrt{\left[r_{E\phi}^2(\phi)-a^2r^2\right]V'_{t0}{}^2(\phi)}}\ ,
\label{SvtnotO4}
\end{align}
and for the action calculated in the Euclidean approach we have
\begin{align}
    S_E=\int d^3x \int_{-\infty}^{\infty}d\tau\, \left[\frac12 \left(\frac{\partial \phi_B}{\partial\tau}\right)^2+
    \frac12 \left(\nabla \phi_B\right)^2+ V(\phi_B,\vec{x})\right]\ ,
\end{align}
where now $\phi_B=\phi_B(\vec{x},\tau)$.
Proceeding as above we get
\begin{align}
    S_E=\int d^3x \int_{-\infty}^{\infty}d\tau\,
    \left\{\frac{V_{t0}(\phi_B)}{3}\nabla\cdot \vec{\omega}_t +\frac19\left[r_{E\phi}^2(\phi_B)-a^2r^2\right]V_{t0}'{}^2(\phi_B)\right\} \ ,
\end{align}
where now $\vec{\omega}_t$, $r_{E\phi}$ and $V_{t0}$ are evaluated at $\phi_B(\vec{x},\tau)$. 
We reproduce the result (\ref{SvtnotO4}) if we transform the $\tau$ integrals into $\phi$ integrals using
\begin{align}
    \int_{-\infty}^{\infty} d\tau\, f[\vec{x},\tau,\phi_B(\vec{x},\tau)] = 
    2\int_0^{\phi_B(\vec{x},0)} d\phi\, \frac{3f[\vec{x},\sqrt{r_{E\phi}^2(\phi)-a^2r^2},\phi]}{\sqrt{\left[r_{E\phi}^2(\phi)-a^2r^2\right] V_{t0}'{}^2(\phi)}}\ ,
\end{align}
with the overall factor of 2 accounting for the fact that the integral in $\tau$ covers twice the field range from $\phi_B(\vec{x},\pm\infty)=0$
to $\phi_B(\vec{x},0)$, with $\phi_B(\vec{x},-\tau)=\phi_B(\vec{x},\tau)$.

We can also obtain a simple expression for the energy of the nucleated bubble (at $\tau=0$), $\phi_B(\vec{x},0)$, which is given by
\begin{align}
    E=  \int d^3x\, \left[\frac12 \left(\frac{\partial \phi_B}{\partial\tau}\right)^2+
    \frac12 (\nabla \phi_B)^2+ V(\phi_B,\vec{x})\right]\ .
\end{align}
Using the previous results we can rewrite this as
\begin{align}
    E= \frac{V_{t0}(\phi)}{3}\int d^3x\, \nabla\cdot\vec{\omega_t} =\frac{V_{t0}(\phi)}{3}\int d\vec{S}\cdot \vec{\omega_t}\ ,
\end{align}
where  it is understood that $\vec{\omega_t}$ is evaluated at $\phi=\phi_B(\vec{x},0)$. Given the asymptotic behaviour assumed for $a$
and $V_{t0}$ we get $E=0$.

%%%%%%%%%%%%%%%%%%%%%%%%%%%%%%%%%%%%%%%%%%%%%%%%%
\subsection{\texorpdfstring{$\bma{O(3)}$}{O(3)} Deformed example with thin-wall limit}
\label{subsec:thinwallO(3)}
%%%%%%%%%%%%%%%%%%%%%%%%%%%%%%%%%%%%%%%%%%%%%%%%%

For this concrete example we take the Euclidean bounce to be
\begin{align}
\phi_B(r,\tau)=\frac{\phi_0}{\phi_0+(1-\phi_0) \exp{[a^2(r)r^2+\tau^2]}}\ .
\end{align}
The $O(4)$-symmetric bounce, for $a(r)=1$, was first presented in \cite{Espinosa:2018hue}.
It describes the decay of the metastable $\phi=0$ vacuum of the potential  
\begin{align}
V_0(\phi)= \phi^2\left[2\phi-3+(1-\phi)^2\log\frac{(1-\phi)^2\phi_0^2}{(1-\phi_0)^2\phi^2}\right]\ ,   
\end{align}
with mass parameters set to 1 for simplicity. The false vacuum is at $\phi=0$ and the true one at $\phi=1$.
The corresponding tunneling potential is 
\begin{align}
    V_{t0}(\phi)=\phi^2(2\phi-3)\, ,
\end{align}
and the tunneling action is given analytically by
\begin{align}
S_0=-\frac{\pi^2}{3}\left[\phi_0+\mathrm{Li}_2\left(\frac{\phi_0}{\phi_0-1}\right)\right] \ .
\label{S0}
\end{align}
The field value $\phi_0$ (with $0<\phi_0<1$) is the end point of the tunneling solution so that we have in fact a family of 
bounces parametrized by $\phi_0$. When $\phi_0\to 1$ these bounces are in the thin-wall regime.

In this example the function $r_{E\phi}^2(\phi)$, defined by (\ref{rE}), is
\begin{align}
    r_{E\phi}^2(\phi)= \log\frac{\phi_0(1-\phi)}{\phi(1-\phi_0)}\ .
\end{align}
Using the general formula (\ref{Vphix}) we obtain the potential 
(up to an additive arbitrary function of $r$) as
\begin{align}
V(\phi,r)= V_0(\phi)+\frac{V_{t0}(\phi)}{3r^2}\frac{d}{dr}\left\{r^3 \left[a(a+r\dot a)-1\right]\right\}
+\frac{V_{t0}'{}^2(\phi)}{18}r^2a^2\left[(a+r\dot a)^2-1\right]\ ,
\end{align}
with $\dot a=da/dr$. Notice that for this $r$-dependent potential, the false and true vacua are still at $\phi=0$ and $\phi=1$ respectively,
without an $r$-dependence.

The energy of the tunneling bubble [given by the $\tau=0$ slice of the bounce, $\phi_B(r,0)$] is 
\begin{align}
    E=\left.
    \frac{4\pi r^3}{3}a(a+\dot{a}r)\left[2\phi_B(r,0)-3\right]\phi^2_B(r,0)
    \right|_{r=0}^{r=\infty}\ ,
\end{align}
which gives $E=0$ given the right asymptotics of $a(r)$ and $\phi_B(r,0)$.

The Euclidean action is
\begin{align}
    S_E&=16\pi \int_0^\infty dr \int_{-\infty}^\infty d\tau\  r^2\tau^2 \phi_B(r,\tau)^2\left[1-\phi_B(r,\tau)\right]^2\nonumber\\
    &+\left.\frac{4\pi}{3} \int_{-\infty}^\infty d\tau\ r^3 \left[2\phi_B(r,\tau)-3\right]\phi_B(r,\tau)^2 a(r)\left[a(r)+\dot{a}(r)r\right]\right|_{r=0}^{r=\infty}
    \ ,
\end{align}
and the boundary term vanishes for appropriate asymptotics of $a(r)$.

In the tunneling potential formalism we get
\begin{align}
\V_t(\phi,r) = \frac{\phi^2(2\phi-3)}{3r^2} \frac{d}{dr}\left[r^3a(a+r\dot a)\right]\ ,
\end{align}
and thus
\begin{align}
    V_t(\phi,r)= \frac{3}{r^3}\int_0^r \V_t(\phi,\bar r)
    \bar r^2d\bar r= \phi^2(2\phi-3) a(r)\left[a(r)+\dot{a}(r)r\right]\ ,
\end{align}
where we used $\lim_{r\to 0}r^3 a(r)\left[a(r)+\dot{a}(r)r\right]=0$.
The corresponding tunneling action can be put in a simple form, with a bulk contribution given by
\begin{align}
S =  16\pi \int_0^\infty dr \int_0^{\phi_B(r,0)}d\phi\ r^2 \phi(1-\phi)\sqrt{\log\frac{(1-\phi)\phi_B(r,0)}{[1-\phi_B(r,0)]\phi}}\ ,
\label{svtex}
\end{align}
so that, in the notation of Sec.~\ref{sec:without_gravity}, we have $\phi_m(r)=0$ and $\phi_M(r)=\phi_B(r,0)$.
The boundary action vanishes as in the Euclidean expression.
 
\begin{figure}[t!]
\begin{center}
\hspace*{-2cm}\includegraphics[width=1.2\textwidth]{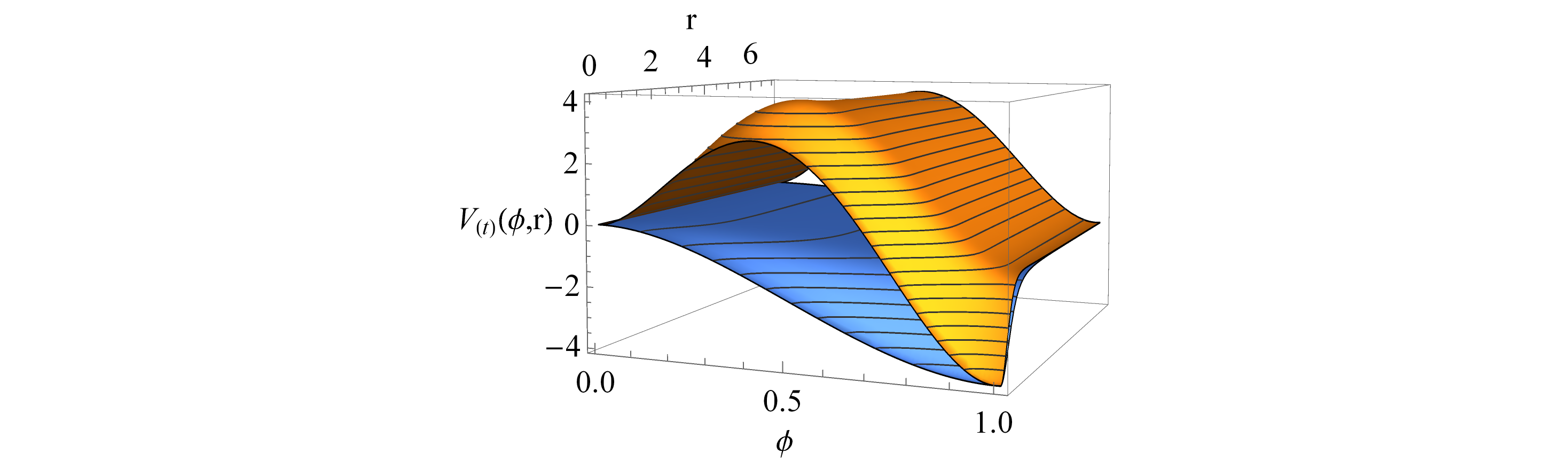}
\end{center}
\caption{ Potential $V(\phi,r)$ and tunneling potential $V_t(\phi,r)$ for the deformed thin-wall example 
of subsection \ref{subsec:thinwallO(3)} with $\alpha=1$ and $\phi_0$ chosen
to give a bubble radius $R\simeq 6$. 
\label{fig:VVtstw}
}
\end{figure}

\begin{figure}[t!]
\begin{center}
\includegraphics[width=0.7\textwidth]{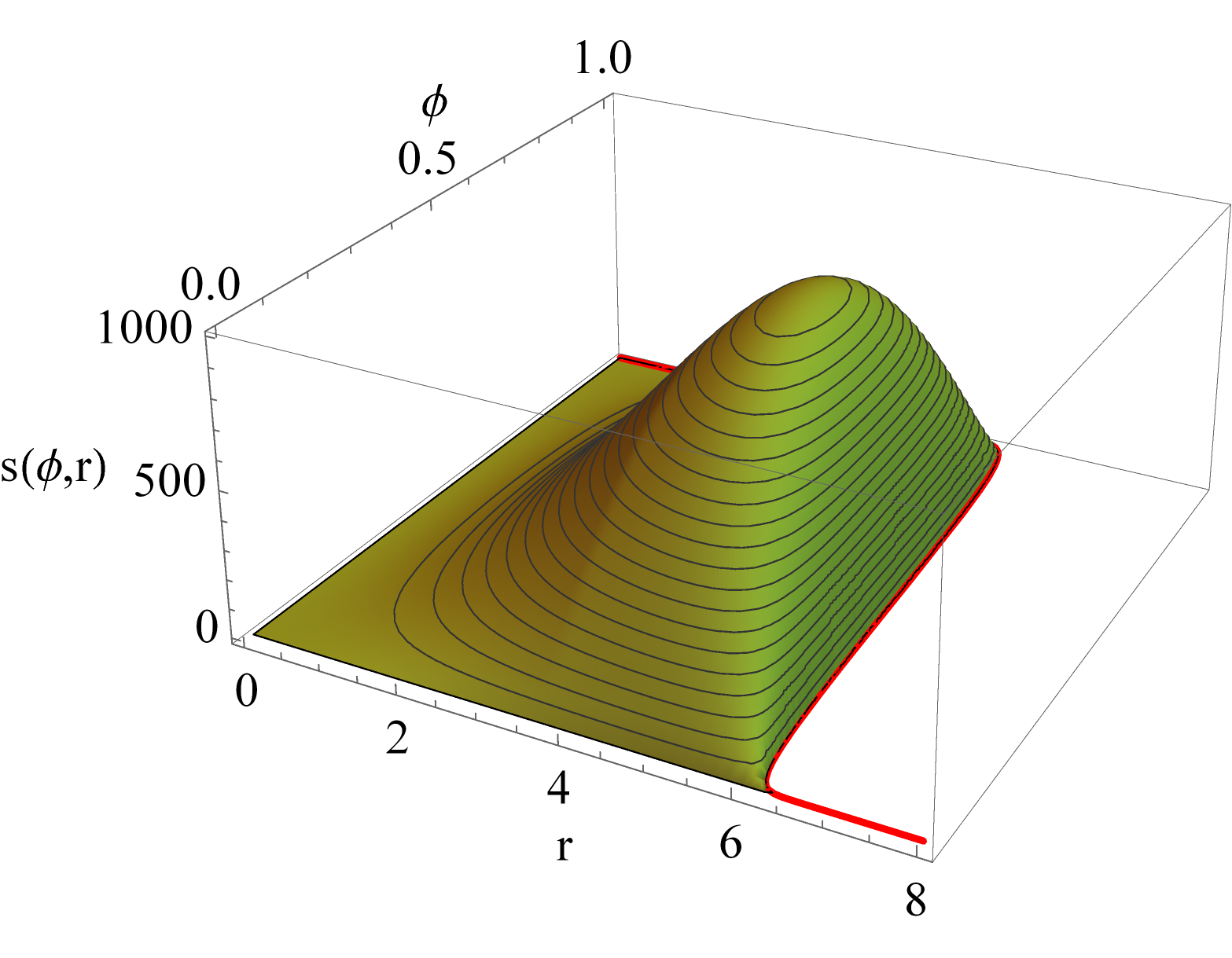}
\end{center}
\caption{Action density $s(\phi,r)$ for the deformed thin-wall example of subsection \ref{subsec:thinwallO(3)} with $\alpha=1$ and $\phi_0$ chosen
to give a bubble radius $R\simeq 6$. The red line shows $\phi_B(r,0)$.
\label{fig:Stw}
}
\end{figure}

For numerical work we choose
\begin{align}
    a(r)=1+\frac{\alpha}{1+r^2}\ ,
\end{align}
where $\alpha$ controls the amount of departure from the $O(4)$-symmetric case ($\alpha=0$).
Figure~\ref{fig:VVtstw} shows the potential $V(\phi,r)$ and tunneling potential $V_t(\phi,r)$ for $\alpha=1$ and $\phi_0$ chosen
very close to $1$ so as to give a bubble radius $R\simeq 6$ (a thin-wall case). Figure~\ref{fig:Stw} gives the action density
$s(\phi,r)$, the integrand in (\ref{svtex}), for the same numerical case. The red line shows the profile of the critical bubble $\phi_B(r,0)$, bounding the domain where $s(\phi,r)$ is defined.

\begin{figure}[t!]
\begin{center}
\hspace*{-2cm}\includegraphics[width=0.5\textwidth]{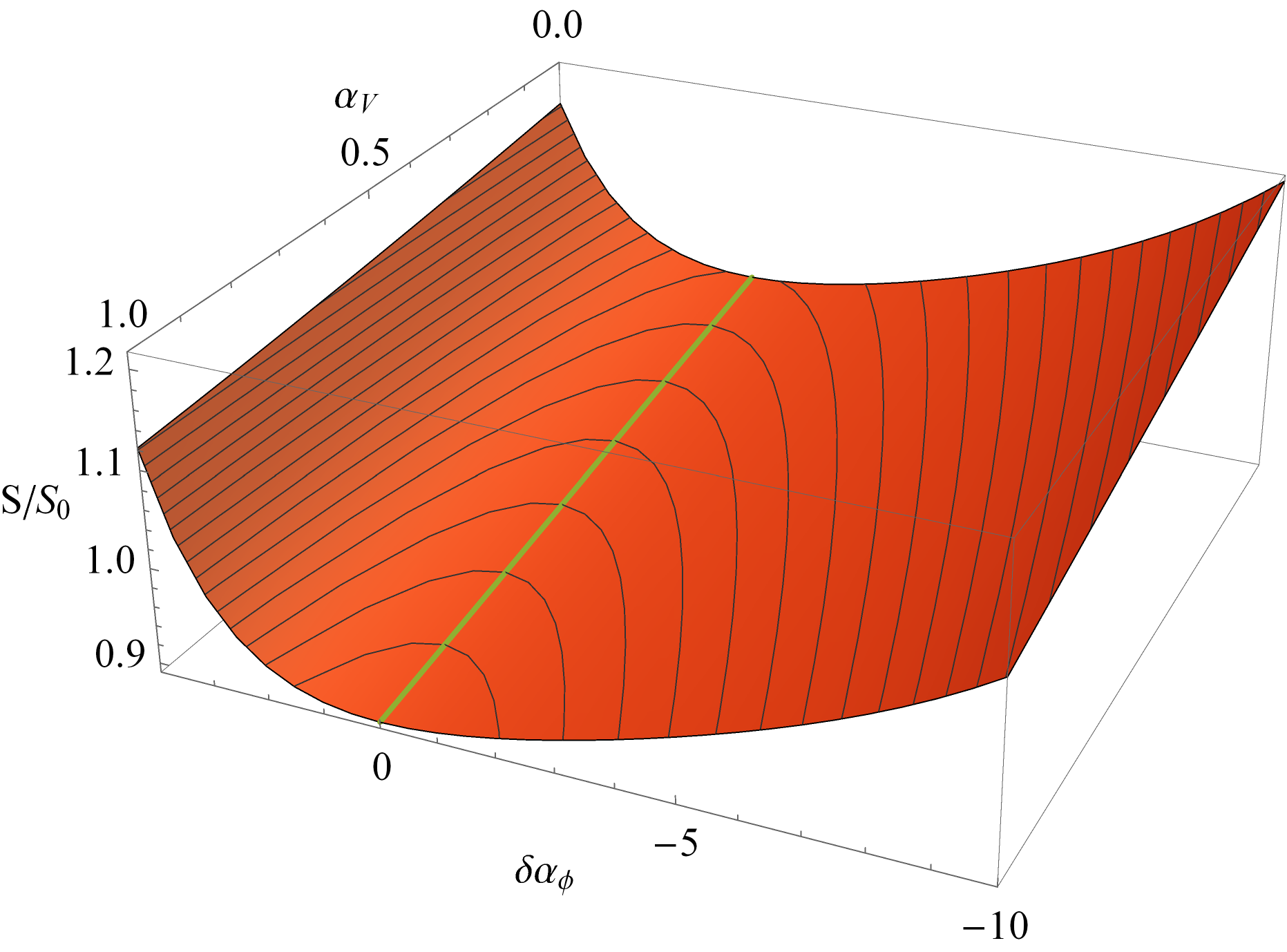}
\end{center}
\caption{ Action (normalized to the $O(4)$ value) for the deformed thin-wall example of subsection \ref{subsec:thinwallO(3)} 
with $\phi_0$ chosen to give a bubble radius $R\simeq 6$, $\alpha=\alpha_V$ in the potential and $\alpha=\alpha_V+\delta\alpha_\phi$ in the 
profile of the deformed bounce.
The green line shows the case with same $\alpha$ value in both functions.
\label{fig:alphaLandscape}
}
\end{figure}

To illustrate how the deformed bounce minimizes the tunneling action we can consider different values of $\alpha$ for the potential ($\alpha_V$) and for the bounce ($\alpha_\phi$). A fixed value of $\alpha_V$ fixes a certain theory/defect and different values of 
$\alpha_\phi$ different field profiles. Figure~\ref{fig:alphaLandscape} shows the action [normalized to the $O(4)$ value in (\ref{S0}) as a function
of $\alpha_V$ for profiles with $\alpha_\phi=\alpha_V+\delta\alpha_\phi$.\footnote{For a proper calculation of the action the bounce has to be further rescaled along its radial direction to ensure
$E=0$ holds.} We see that, for a given $\alpha_V$ the action is minimized
by the $\delta\alpha_\phi=0$ profile (with equal $\alpha$'s for potential and bounce, green line) while the action decreases with increasing $\alpha_V$
along the "theory/defect axis". 

Needless to say, following our procedure many other analytical examples can be constructed from other $O(4)$ symmetric bounces,
e.g. the Fubini bounce for $V_0(\phi)=-\lambda\phi^4/4$ or other examples in the literature (see e.g. \cite{Espinosa:2023oml}).

%%%%%%%%%%%%%%%%%%%%%%%%%%%%%%%%%%%%%%%%%%%%%%%%%%
\section{Discussion and conclusions}
\label{sec:conclusion}
%%%%%%%%%%%%%%%%%%%%%%%%%%%%%%%%%%%%%%%%%%%%%%%%%%

In this paper, we constructed a positive-definite tunneling action,  the $S[V_t]$ given in (\ref{eq:action_without_gravity}), for decays with $O(3)$ symmetry only (that is, rotationally invariant in real space but not in Euclidean spacetime).  The dynamical variable in this approach is the tunneling potential $V_t(\phi, r)$, which resides in $(\phi, r)$-space. The method is an extension of the tunneling potential approach, which was originally applied to $O(4)$-symmetric tunneling. The main motivation of this generalization is tunneling around impurities. Examples of such impurities include monopoles and non-rotating black holes with $O(3)$ symmetry. Such an extension is needed because tunneling around the relevant bounces break $O(4)$ symmetry.

In the derivation, we first use the monotonicity of the bounce in the Euclidean time direction to exchange the roles of the field variable and the Euclidean time coordinate, rewriting the original configuration $\phi(\tau,r)$ in terms of its inverse $\tau(\phi,r)$. The three-dimensional radial coordinate $r$ is kept as an independent variable throughout this change of variables. We then pass to the Hamiltonian form and perform a canonical transformation generated by (\ref{eq:generating_function}), which replaces the original canonical variables by $Q=-\dot{\tau}$ and \(P\). After eliminating \(Q\), the action can be written solely in terms of \(P\), and finally in terms of the generalized tunneling potential through \(V_t=3P/(4\pi r^3)\). This procedure closely parallels the canonical-transformation derivation of the O(4)-symmetric tunneling potential, but it only assumes $O(3)$ symmetry rather than $O(4)$ symmetry.

While the dynamical variable $V_t(\phi)$ of the $O(4)$ tunneling potential method resides in $\phi$-space, the dynamical variable $V_t(\phi,r)$ in the generalized formulation resides in $(\phi,r)$-space. This reflects the fact that the derivation assumes only $O(3)$ symmetry and does not impose $O(4)$ symmetry. We only interchanged the roles of the dynamical variable $\phi$ and the time coordinate $\tau$, not those of the spatial coordinate $r$. 

Since the steps leading to the new functional $S[V_t]$ are based only on the canonical transformation, its on-shell value agrees with that of the Euclidean bounce action in the $\phi(\tau,r)$ formulation. The square-root structure also has a simple interpretation in terms of the original bounce variables, with the endpoint of the $\phi$-integration identified with the $\tau=0$ slice of the bounce, $\phi_B(0,r)$, and the lowest with the false vacuum.

The tunneling action obtained, Eq.~(\ref{eq:action_without_gravity}), is manifestly positive definite, as it is written in square-root form. This attractive property is shared with the tunneling potential result for the $O(4)$ case. As shown in Sec.~\ref{subsec:without_gravity_O(4)limit}, one can establish the relation between the $O(3)$ and $O(4)$ tunneling potentials. The $O(4)$ limit is realized via the $r$-independence of the dynamical variable $V_t(\phi,r)$, and the action and EoM reduce to those of the $O(4)$ tunneling-potential method.

This extended formulation and its possible extensions open several avenues for future work.
A natural direction is to extend the present construction to configurations with less symmetry than $O(3)$ symmetry, or even without any symmetry. Such situations are relevant, for example, to tunneling in the presence of non-spherical impurities, such as cosmic strings and domain walls, where more than one spatial coordinate has to be kept as an independent variable. In such cases, the structure of the canonical transformation and the treatment of boundary terms  becomes more involved than in the $O(3)$-symmetric case studied here. Such an extension may also open up the possibility of formulating the one-loop determinant within the tunneling potential formulation, since the fluctuation modes around the bounce are not restricted to preserve the symmetry of the background. We leave a systematic construction of this lower-symmetry formulation for future work. 

Another important extension is the analysis of tunneling with fully dynamical gravity. In Ref.~\cite{Espinosa:2018voj}, the $O(4)$ tunneling potential method with Minkowski spacetime, de Sitter spacetime and Anti-de Sitter spacetime was formulated. The method is also efficient for tunneling in the presence of gravity. The most notable feature in applications is that Coleman-de Luccia bounce and Hawking-Moss bounce can be compared within the same calculation since one does not need to care about boundary conditions.
In these situations, we can assume $O(4)$ symmetry. However, there are some interesting tunneling phenomena including gravity without $O(4)$ symmetry. A typical example is vacuum decay around black holes, which can strongly enhance the decay rate. In the presence of black holes, we can assume $O(3)$ symmetry with non-rotating black holes and $O(2)$ symmetry with rotating black holes. The extension of the positive-definite tunneling action  (\ref{eq:action_without_gravity}) is potentially effective for the analysis of tunneling in these situations.

Finally, it would be interesting to see how numerical implementation works with the tunneling action introduced in this paper.
We leave this for future work.

%%%%%%%%%%%%%%%%%%%%%%%%%%%%%%%%%%%%%%%%%%%%%%%%%%
\section*{Acknowledgements}
%%%%%%%%%%%%%%%%%%%%%%%%%%%%%%%%%%%%%%%%%%%%%%%%%%

J.R.E acknowledges CERN TH Department for hospitality while this research was being carried out. The work of J.R.E. has been supported by the IFT Centro de Excelencia Severo Ochoa CEX2020-001007-S and by PID2022-142545NB-C22 funded by MCIN/AEI/10.13039/ 501100011033 and by “ERDF A way of making Europe”. 
The work of R.J. is supported by JSPS KAKENHI Grant Number 23K19048 and 24K07013.
TK is supported by the Deutsche Forschungsgemeinschaft (DFG, German Research Foundation) under Germany’s Excellence Strategy – EXC 2121 “Quantum Universe” – 390833306.
T.~M.~was supported by JSPS KAKENHI Grant Number JP23KJ1543 and in part by the 2025 Osaka Metropolitan University (OMU) Strategic Research Promotion Project (Development of International Research Hubs).

%%%%%%%%%%%%%%%%%%%%%%%%%%%%%%%%%%%%%%%%%%%%%%%%%%
\bibliographystyle{JHEP}
\bibliography{main}
%%%%%%%%%%%%%%%%%%%%%%%%%%%%%%%%%%%%%%%%%%%%%%%%%%

%%%%%%%%%%%%%%%%%%%%%%%%%%%%%%%%%%%%%%%%%%%%%%%%%%
\end{document}